%% file: spaceconf.tex
\documentclass[10pt, conference, compsocconf]{IEEEtran}
\usepackage{url}
\usepackage{amsmath,amssymb,amsfonts}

\usepackage{tikz}
\usetikzlibrary{shapes.geometric, arrows, shapes,backgrounds,arrows,calc,positioning}
\tikzstyle{process} = [rectangle, minimum width=2.5em, minimum height=2em, text centered, draw=black]
\tikzstyle{arrow} = [thick,->,>=stealth]
\usepackage{tablefootnote}

\usepackage{xspace}
\usepackage{adjustbox}
\usepackage{pifont}% http://ctan.org/pkg/pifont
\newcommand{\xmark}{\ding{55}}%
\usepackage{tabularx}
\IEEEoverridecommandlockouts 
\begin{document}
\title{Modular Verification of Autonomous Space Robotics
\thanks{This work is supported by grant EP/R026092 (FAIR-SPACE Hub) through UKRI under the Industry Strategic Challenge Fund (ISCF) for Robotics and AI Hubs in Extreme and Hazardous Environments.}}


\author{\IEEEauthorblockN{Marie Farrell, Rafael C. Cardoso, Louise A. Dennis, Clare Dixon, Michael Fisher,\\ Georgios Kourtis, Alexei Lisitsa, Matt Luckcuck and Matt Webster}\\

\IEEEauthorblockA{Department of Computer Science\\
University of Liverpool\\
\tt\footnotesize{marie.farrell@liverpool.ac.uk}}
%\and
%\IEEEauthorblockN{Authors Name/s per 2nd Affiliation (Author)}
%\IEEEauthorblockA{line 1 (of Affiliation): dept. name of organization\\
%line 2: name of organization, acronyms acceptable\\
%line 3: City, Country\\
%line 4: Email: name@xyz.com}
}


%Telephone: (800) 555--1212, Fax: (888) 555--1212}


\maketitle

\begin{abstract}
%Software engineering of autonomous robotic systems deployed in space is a challenging task that is complicated by unreliable communications, as well as a hazardous and unpredictable environment. 
Ensuring that autonomous space robot control software behaves as it should is crucial, particularly as software failure in space often equates to mission failure and could potentially endanger nearby astronauts and costly equipment. To minimise mission failure caused by software errors, we can utilise a variety of tools and techniques to verify that the software behaves as intended. In particular, distinct nodes in a robotic system often require different verification techniques to ensure that they behave as expected. %Ensuring whole system consistency when individual nodes are verified using a heterogeneous techniques and formalisms is difficult. 
This paper introduces a method for integrating the various verification techniques that are applied to robotic software, via a First-Order Logic (FOL) specification that captures each node's assumptions and guarantees. These FOL specifications are then used to guide the verification of the individual nodes, be it by testing or the use of a formal method. We also outline a way of measuring our confidence in the verification of the entire system in terms of the verification techniques used.
%In this paper, we provide a simple example of the former and we also discuss a way of defining a confidence metric for the verification associated with each individual node
\end{abstract}

\begin{IEEEkeywords}
formal methods; heterogeneous verification; autonomous space robotics
\end{IEEEkeywords}

\IEEEpeerreviewmaketitle

\section{Introduction}
%Autonomous robotic systems are being used more frequently in safety-critical scenarios (e.g. \cite{Hastie:18}, \cite{Bogue:11}) so it is no surprise that they will be involved in space missions in the near future~\cite{Wilcox:92}.  % modern robotic systems become more autonomous and complex, and are deployed alongside humans.

%Robotic software is often developed using node-based software frameworks %, such as ROS\footnote{\url{http://www.ros.org/}}, 
%to facilitate the decoupling of software modules. These frameworks share abstract concepts~\cite{Shakhimardanov2010}, in that systems are composed of communicating nodes. 

Robotic systems combine many hardware and software components, usually represented as node-based architectures.
Each node in a robotic system may require different verification techniques, ranging from software testing to formal methods. In fact, \textit{integrating} (formal and non-formal) verification techniques is crucial for the robotics domain~\cite{Farrell2018}. %, particularly as these robots are deployed in hazardous environments potentially far away from any human intervention, such as in space. 
Verification should be carried out using the most suitable technique or formalism for each node. However, linking heterogeneous verification results of individual nodes is difficult and the current state-of-the-art for robotic software development does not provide an easy way of achieving this. 

%Our proposed method is to construct a high-level First-Order Logic (FOL) specification of each node in the system. The FOL specification describes the expected input, required output, and Assume-Guarantee \cite{Jones83} conditions for each node. 

%Our method can be used Top-Down, to guide the system's development from abstract specification to concrete implementation, via verification; or Bottom-Up, to check the  consistency of existing verification technique. In this paper we show a Top-Down example of applying our method. It enables developers to choose the most suitable verification technique for each node, but also to link the conditions being verified across the whole system.
In  Fig. \ref{fig:nodespecs}, we consider a simple space robotic system: a planetary rover undertaking a remote inspection task. Here, we have nodes representing the \textbf{Vision} system, a \textbf{Planner} that returns a set of potential plans between the current location and the next point to inspect, an autonomous \textbf{Plan Reasoning Agent} that selects a plan, and a \textbf{Hardware Interface} that sends commands to the rover's actuators. 

As illustrated by Fig. \ref{fig:nodespecs}, we could use logical specifications (e.g. temporal logic), model-based specifications (e.g. Event-B or Z), or algebraic specifications (e.g. CSP or CASL) amongst others to specify the nodes in a robotic system. Each of these formalisms offers its own range of benefits, and each tends to suit the verification of particular types of behaviour. However, in some cases we may only have access to the black-box or white-box implementation of a node and so, we must use (simulation-based) testing techniques for verification. 

Our approach facilitates the use of heterogeneous verification techniques for the nodes in a robotic system. We achieve this by specifying Assume-Guarantee \cite{Jones83} properties in FOL, as high-level node specifications, and we employ temporal logic for reasoning about the combination of these FOL specifications. Thus, we attach the assumptions ($\mathcal{A}(\overline{i})$) and guarantees ($\mathcal{G}(\overline{o})$) to individual nodes (shown in Fig. \ref{fig:nodespecs}). This abstract specification can be seen as a logical prototype for individual nodes and thus the entire robotic system. %These FOL specifications are then used to guide the verification of each node by encoding the assumptions and guarantees (as, for example, test cases, assertions, or formal properties).
\section{FOL Assume-Guarantee Specifications}
\input{agnodes}
%No single verification approach suits every node in a robotic system~\cite{Farrell2018}. For example, hardware interfaces or planners may be amenable to formal verification, whereas, sensor systems may require software testing. % or simulation-based testing. 

%

%Specifically, we use \emph{typed} FOL, potentially with the addition of algebraic operators, to specify the assumptions and guarantees. 
For each node, $N$, we specify $\mathcal{A}_N(\overline{i}_N)$ and $\mathcal{G}_N(\overline{o}_N)$ where $\overline{i}_N$ is a variable representing the input to the node, $\overline{o}_N$ is a variable representing the output from the node, and $\mathcal{A}_N(\overline{i}_N)$ and $\mathcal{G}_N(\overline{o}_N)$ are FOL formulae describing the assumptions and guarantees, respectively, of this node.

\noindent Each individual node, $N$, obeys the following implication\\
\centerline{$\forall \overline{i}_N, \overline{o}_N \cdot  \mathcal{A}_N(\overline{i}_N) \Rightarrow\ \lozenge \mathcal{G}_N(\overline{o}_N)$}

\noindent where `$\lozenge$' is LTL's~\cite{Pnueli77temporal} ``eventually'' operator. So, this implication means that if the assumptions, $\mathcal{A}_N(\overline{i}_N)$, hold then  \textit{eventually} the guarantee, $\mathcal{G}_N(\overline{o}_N)$, will hold. Note that our use of temporal operators here is motivated by the temporal nature of robotic systems and will be of use in later extensions of this work.

Consider the autonomous \textbf{Plan Reasoning Agent} in Fig. \ref{fig:nodespecs}, we can specify the following simple assumption, $\mathcal{A}_3(\overline{i}_3)$:

\vspace{-10pt}
\small
\begin{align*}
\mathcal{A}_3(\overline{i}_3) = & \ \forall p \cdot p \in PlanSet  \Rightarrow goal \in p
\end{align*}

\normalsize
\noindent which ensures that every plan that is returned by the \textbf{Planner} contains the $goal$ location. Then,
we might specify the guarantee that the agent chooses the shortest $plan$ as follows:

\vspace{-10pt}
\small
\begin{align*} 
\mathcal{G}_3(\overline{o}_3) = & \ plan \in PlanSet \\ 
& \ \ \land \forall p \cdot p \in PlanSet \land  p \neq plan \\
& \ \ \ \ \Rightarrow length(plan) \leq length(p) 
\end{align*}\normalsize

Once the FOL assumption and guarantee are specified, then we use these high-level specifications as properties to be verified of the individual nodes. For the autonomous \textbf{Plan Reasoning Agent}, we can use a number of techniques for verifying that it meets its associated FOL specification. For example, we can specify the node using the \textsc{Gwendolen} agent programming language and then use the AJPF model-checker to verify that it behaves as specified \cite{Dennis2012}.

Nodes in a modular robotic architecture are linked
together and transmit data between them so long as their types/requirements match. Similarly, we can
compose the assume-guarantee specifications of individual nodes in a number of ways and we are working towards a calculus of inference rules that capture this behaviour. To this end, we are developing rules for sequentially composing, joining, branching and looping between nodes.

%, the simplest being
%a sequential composition. The basic way to describe
%such structures is to first have the specification capture \emph{all} of
%the input and output streams and then to describe how these are
%combined in the appropriate inference rules. 
%The proof rule for such linkage is as expected:
%$$
%\begin{array}{l}
%\forall \overline{i}_1,\overline{o}_1.\ \mathcal{A}_1(\overline{i}_1)\ \Rightarrow\ \lozenge \mathcal{G}_1(\overline{o}_1)\\
%\forall\overline{i}_2,\overline{o}_2.\ \mathcal{A}_2(\overline{i}_2)\ \Rightarrow\ \lozenge \mathcal{G}_2(\overline{o}_2)\\
%\overline{o_1} = \overline{i}_2\\
%\vdash\ \forall\overline{i}_2,\overline{o}_1.\ \mathcal{G}_1(\overline{o}_1)\ \Rightarrow\ \mathcal{A}_2(\overline{i}_2)\\ \hline
%\forall\overline{i}_1,\overline{o}_2.\ \mathcal{A}_1(\overline{i}_1)\ \Rightarrow\ \lozenge\mathcal{G}_2(\overline{o}_2)
%\end{array}
%\hspace{30pt} \text{(PR1)}
%$$
%Intuitively, this states that if two nodes are sequentially composed with the output of the first equal to the input of the second and the guarantee of the first implies the assumption of the second then, we can deduce that the assumption of the first node implies that the guarantee of the second will ``eventually'' hold.

\section{Measuring Confidence in Verification}
A key question is how using these different verification techniques affects our confidence in the verification of the whole system. One might think that a formal proof of correctness corresponds to a higher level of confidence than simple testing methods (especially over unbounded environments). However, formal verification is usually only feasible on an abstraction of the system whereas testing can be carried out on the implemented code. Therefore, it is our view that we achieve higher levels of confidence in verification when multiple verification methods have been employed for each node in the system \cite{webster2016corroborative}. 
\begin{table}
\centering
\begin{tabular}{|m{2cm}|m{1.5cm}|m{1.5cm}|m{1.5cm}|}
\hline
 & Testing & Simulation-Based Testing & Formal Methods \\
\hline \hline
\textbf{Vision} & \checkmark & \xmark& \xmark \\
\hline
\textbf{Planner}  &\checkmark & \checkmark& \checkmark \\
\hline
\textbf{Plan Reasoning Agent} &\checkmark & \checkmark& \checkmark \\\hline
\textbf{Hardware Interface}  &\checkmark & \checkmark& \xmark \\\hline
\end{tabular}
\caption{Verification techniques applied to each node.}
\label{table:confidence}
\vspace{-20pt}
\end{table}

We have broadly partitioned current verification techniques into three categories: testing, simulation-based testing and formal methods. We have determined which of these techniques might be employed for each node in our simple example as shown in Table \ref{table:confidence}. We then provide a score for our level of confidence in the verification of the whole system as $9/12$, resulting in a confidence measure of $75\%$. Examining how this metric can be calculated for more complex systems with loops is a future direction for this work.

%Note that this approach is related to but distinct from the Safety Integrity Levels (SILs)~\cite{IEC61508} used for the certification of safety-critical systems.

 %We  specifically target the verification approach used and the evidence that we provide could help in obtaining appropriate SIL certification for a particular system. %The combination of components that are certified to different SILs is an open challenge, being tackled by the domain of mixed criticality research~\cite{Burns2013}.

%Based on this, we can see that our \textbf{Plan Reasoning Agent} meets CL3, however, the black-box \textbf{Vision} node in Fig. \ref{fig:nodespecs} is likely to be CL1. When calculating whole system confidence, one approach might take the average of the confidence levels of the nodes whereas another might be cautious and regard the confidence in the verification as the minimum of that of the individual nodes. We are currently exploring a different ways to calculate entire system confidence which may include weighting the nodes so that the most mission/safety critical ones are assigned the highest importance when measuring whole system confidence.

\section{Conclusions}
When verifying complex robotic systems, it is clear that no single verification technique is suitable for every node in the system~\cite{Farrell2018} and so a logical framework that allows us to integrate the results from distinct verification techniques is needed. We have outlined an initial approach to specifying assumptions and guarantees using FOL for individual nodes in robotic systems and we have used a simple, illustrative example of a planetary rover to convey our approach. Once the FOL specifications have been constructed, they are then used to guide the more detailed verification of each node. Furthermore, we introduce the notion of confidence in verification techniques and provide a broad categorisation. 

Our current work involves developing a calculus for reasoning about and combining the Assume-Guarantee specifications of individual nodes. In the future, we plan to provide tool support for this and to evaluate it using a set of more complex robotic space missions. We also intend to further investigate the suitability of the confidence levels that we have proposed in this paper.

\bibliographystyle{abbrv}
\bibliography{bibliography}
\end{document}

%% file: agnodes.tex
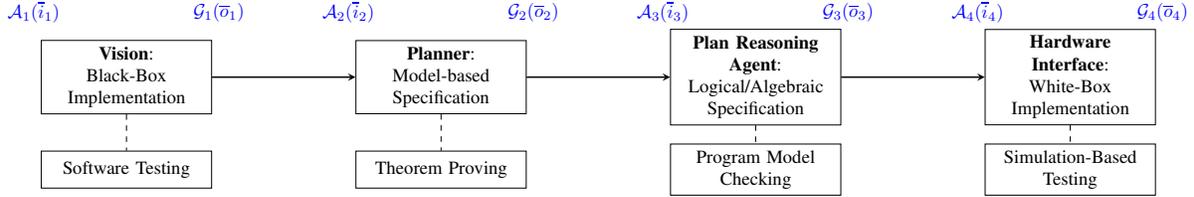
\begin{figure*}[!ht]
\centering
\scalebox{0.7}{
\begin{tikzpicture}[node distance=2em]

\node (node1) [process, xshift = -50em]{\parbox[t][][t]{3cm}{\centering{\textbf{Vision}:}\\\centering{ Black-Box Implementation}}};
\node (A1)[left of = node1, xshift = -3em, yshift = 3.5em]{\color{blue}$\mathcal{A}_1(\overline{i}_1)$};
\node (G1)[right of = node1, xshift = 3em, yshift = 3.5em]{\color{blue}$\mathcal{G}_1(\overline{o}_1)$};

\node (node2) [process, right of = node1, xshift = 15em]{\parbox[t][][t]{3cm}{\centering{\textbf{Planner}:}\\ \centering{ Model-based Specification}}};
\node (A1)[left of = node2, xshift = -3em, yshift = 3.5em]{\color{blue}$\mathcal{A}_2(\overline{i}_2)$};
\node (G1)[right of = node2, xshift = 3em, yshift = 3.5em]{\color{blue}$\mathcal{G}_2(\overline{o}_2)$};

\node (node3) [process, right of = node2, xshift = 15em]{\parbox[t][][t]{3cm}{\centering \textbf{Plan Reasoning Agent}:\\ Logical/Algebraic Specification}};
\node (A1)[left of = node3, xshift = -3em, yshift = 3.5em]{\color{blue}$\mathcal{A}_3(\overline{i}_3)$};
\node (G1)[right of = node3, xshift = 3em, yshift = 3.5em]{\color{blue}$\mathcal{G}_3(\overline{o}_3)$};

\node (node4) [process, right of = node3, xshift = 15em]{\parbox[t][][t]{3cm}{\centering{\textbf{Hardware Interface}:}\\\centering{ White-Box Implementation}}};
\node (A1)[left of = node4, xshift = -3em, yshift = 3.5em]{\color{blue}$\mathcal{A}_4(\overline{i}_4)$};
\node (G1)[right of = node4, xshift = 3em, yshift = 3.5em]{\color{blue}$\mathcal{G}_4(\overline{o}_4)$};

%\node(topdots)[right of = node4, xshift = 10em]{$\ldots$};

\node (S13) [process, below of = node1, yshift = -3em]{\parbox[t][][t]{3cm}{\centering{Software Testing}}};

\node (S14) [process, below of = node2, yshift = -3em]{\parbox[t][][t]{3cm}{\centering{Theorem Proving}}};

\node (S15) [process, below of = node3, yshift = -3em]{\parbox[t][][t]{3cm}{\centering{Program Model Checking}}};

\node (S16) [process, below of = node4, yshift = -3em]{\parbox[t][][t]{3cm}{\centering{Simulation-Based Testing}}};

\draw [arrow] (node1) -- node[above]{}(node2);
\draw [arrow] (node2) --  node[above]{}(node3);
\draw [arrow] (node3) -- node[above]{}(node4);
%\draw [arrow] (node4) -- node[above]{}(topdots);
%\draw [arrow, dashed] (S13) -- node[above]{}(S14);
%\draw [arrow, dashed] (S14) -- node[above]{}(S15);
%\draw [arrow, dashed] (S15) -- node[above]{}(S16);

\draw [dashed ] (node1)  -- (S13);
\draw [dashed] (node2)  -- (S14);
\draw [dashed ] (node3)  -- (S15);
\draw [dashed ] (node4)  -- (S16);

\end{tikzpicture}

}
\caption{We specify the Assume-Guarantee properties for each node (denoted by $\mathcal{A}(\overline{i})$ and $\mathcal{G}(\overline{o})$ respectively). These are then used to guide the verification approach applied to each node, denoted by dashed lines, such as software testing for a black-box implementation of the \textbf{Vision} node. The solid arrows represent data flow between nodes and that the assumptions of the next node should follow from the guarantee of the previous.}
\label{fig:nodespecs}
\vspace{-1em}
\end{figure*}